\documentclass[prl,twocolumn]{revtex4}
\usepackage{graphicx}
\usepackage{amssymb,amsmath,bm}
\usepackage{enumerate}

\newcommand{\be}{\begin{eqnarray}}
\newcommand{\ee}{\end{eqnarray}}

\begin{document}

\title{Decaying turbulence and developing chaotic attractors}

\author{A. Bershadskii}

\affiliation{
ICAR, P.O. Box 31155, Jerusalem 91000, Israel
}

\begin{abstract}

Competition between two main attractors of the distributed chaos, one associated  with translational symmetry (homogeneity) and another  associated  with rotational symmetry (isotropy), has been studied in freely decaying turbulence. It is shown that, unlike the case of statistically stationary homogeneous isotropic turbulence, the attractor associated with rotational symmetry (and controlled by Loitsyanskii integral) 
can dominate turbulent local dynamics in an {\it intermediate} stage of the decay, because the attractor associated with translational symmetry (and controlled by Birkhoff-Saffman integral) is still not developed enough. The DNS data have been used in order to support this conclusion. 
\end{abstract}

\maketitle

\section{Introduction}

Freely decaying turbulence becomes similar to isotropic homogeneous turbulence after going through certain stages of development. It can be clear seen on time dynamics of the skewness of the velocity derivatives
$$
S= \frac{\left\langle\left(\frac{\partial u_i}{\partial x_i}\right)^3\right\rangle}{\left\langle\left(\frac{\partial u_i}{\partial x_i}\right)^2\right\rangle^{3/2}} \eqno{(1)}.
$$

For isotropic homogeneous turbulence, the skewness characterises the
rate of vorticity production by vortex stretching \cite{my}.
At relatively small turbulent Reynolds numbers $Re_{\lambda}$ the skewness $S \simeq -0.5$ for the statistically stationary isotropic homogeneous turbulence. Figure 1 shows time dynamics of the skewness in a freely decaying turbulence. The data were taken from site Ref. \cite{torr}. In this site the results of a direct numerical simulation (DNS) of decaying turbulence in a triply periodic box at $512^3$ resolution, performed by A.A.Wray \cite{wr}, are presented. The DNS was started from an uncorrelated random field at $Re_{\lambda} \simeq 950$. When it reached $Re_{\lambda} = 62$, the DNS had resolution $k_{max}\eta=1.4$ (where $\eta$ is the Kolmogorov microscale).

One can see that for $t > 2$ (in the DNS units) the skewness approaches the value -0.5. The dashed straight line in Fig. 1 corresponds to the stationary isotropic homogeneous value $S \simeq - 0.52$ observed in a DNS at $38< Re_{\lambda } < 70$  Ref. \cite{gfn} (Table II).

\begin{figure} 
\begin{center}
\includegraphics[width=8cm \vspace{-0.6cm}]{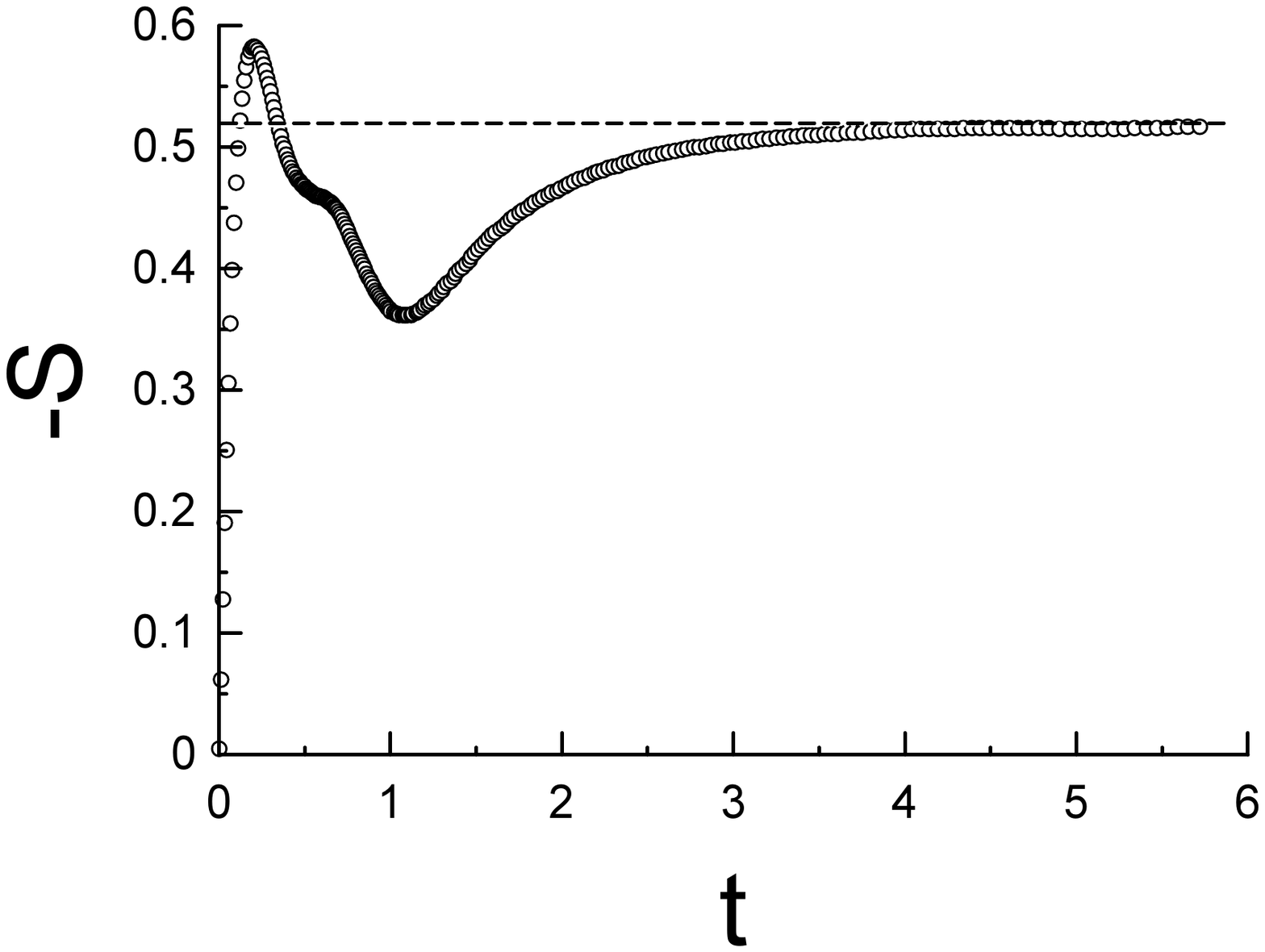}\vspace{-4cm}
\caption{\label{fig1} Skewness of a freely decaying turbulence (in the DNS units). The DNS data \cite{wr} were taken from site Ref. \cite{torr}. The dashed straight line corresponds to the stationary isotropic homogeneous value at $38< Re_{\lambda } < 70$ from DNS Ref. \cite{gfn} (Table II).}
\end{center}
\end{figure}
 Distributed chaos in the statistically stationary isotropic and homogeneous turbulence has two main attractors associated  with the two main space symmetries: translational (homogeneity) and rotational (isotropy) \cite{b1}. Let us denote these attractors as h-attractor and i-attractor correspondingly. Due to the Noether's theorem these attractors are associated also with two main conservation laws: conservation of linear and of angular momentum \cite{ll2}, and through these conservation laws with the Birkhoff-Saffman ($I_2$) and Loitsyanskii ($I_4$) invariants \cite{my},\cite{saf},\cite{dav1}
$$   
I_n = \int r^{n-2} \langle {\bf u} ({\bf x},t) \cdot  {\bf u} ({\bf x} + {\bf r},t) \rangle d{\bf r}  \eqno{(2)}
$$  
 
Let us recall that the invariants which (due to the Noether's theorem) are consequences of the space symmetries are compatible with viscosity dissipation \cite{my},\cite{saf}. Therefore, unlike the invariant (energy) associated with time translational symmetry, these invariants do not demand 
large Reynolds numbers in order to be applicable, and they can be well applied to the near dissipation range of scales \cite{falc},\cite{b2} and to the decaying turbulence with relatively small Reynolds numbers. 
 
  The h- and i-attractors have different sets of initial conditions which eventually approach
each attractor (so called basin of attraction). It is known that basin of attraction of the i-attractor is small and thin in comparison with the h-attractor. Because of this the Birkhoff-Saffman integral $I_2$ dominates the distributed chaos dynamics in the statistically {\it stationary} isotropic and homogeneous turbulence \cite{b1}. But just because of this the i-attractor approaches its developed state earlier than the h-attractor in the {\it decaying} turbulence. The i-attractor needs in less time to involve its small and thin basin of attraction 
than the h-attractor with its large basin of attraction. Therefore, we can expect that in the decaying turbulence there exists an {\it intermediate} range of times where the i-attractor dominates chaotic dynamics just because the h-attractor is still not developed enough. With 
advance of the h-attractor development the competition depends on stability of the i-attractor both to the instant perturbations and to noise. If the energy spectrum of the initial data is chosen (in DNS, for instance) proportional to $k^4$ for small $k$, then the basin of attraction of the i-attractor has certain preference \cite{saf},\cite{dav1},\cite{saf2},\cite{dav2}, that results in an additional delay of the h-attractor development in comparison with i-attractor. 

  It is difficult to distinguish between the Birkhoff-Saffman (h-attractor) and Loitsyanskii (i-attractor) regimes in the decaying turbulence using data on decay of the global variables such as total energy. The energy spectra can be more informative for this purpose, because they provide a broad information about local dynamics in the wavenumbers space at each time point.   
  \begin{figure}
\begin{center}
\includegraphics[width=8cm \vspace{-0.2cm}]{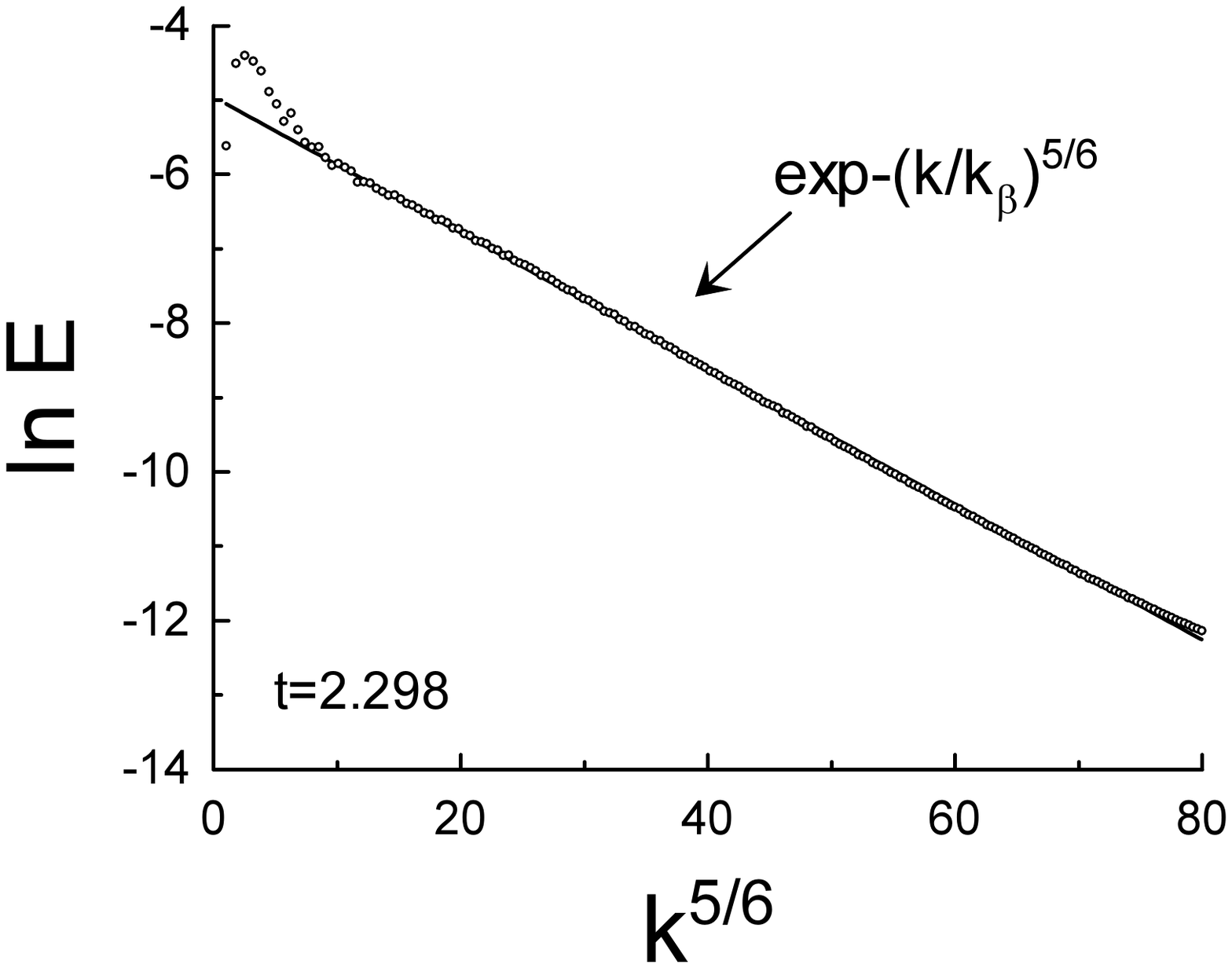}\vspace{-3.8cm}
\caption{\label{fig2} For the same DNS as in Fig. 1 but for logarithm of the 3D velocity power spectrum at $t=2.298$ as function of $k^{5/6}$ (in the DNS units). The solid straight line corresponds to the Eq. (3) with $\beta=5/6$. }
\end{center}
\end{figure}

\begin{figure}
\begin{center}
\includegraphics[width=8cm \vspace{-0.1cm}]{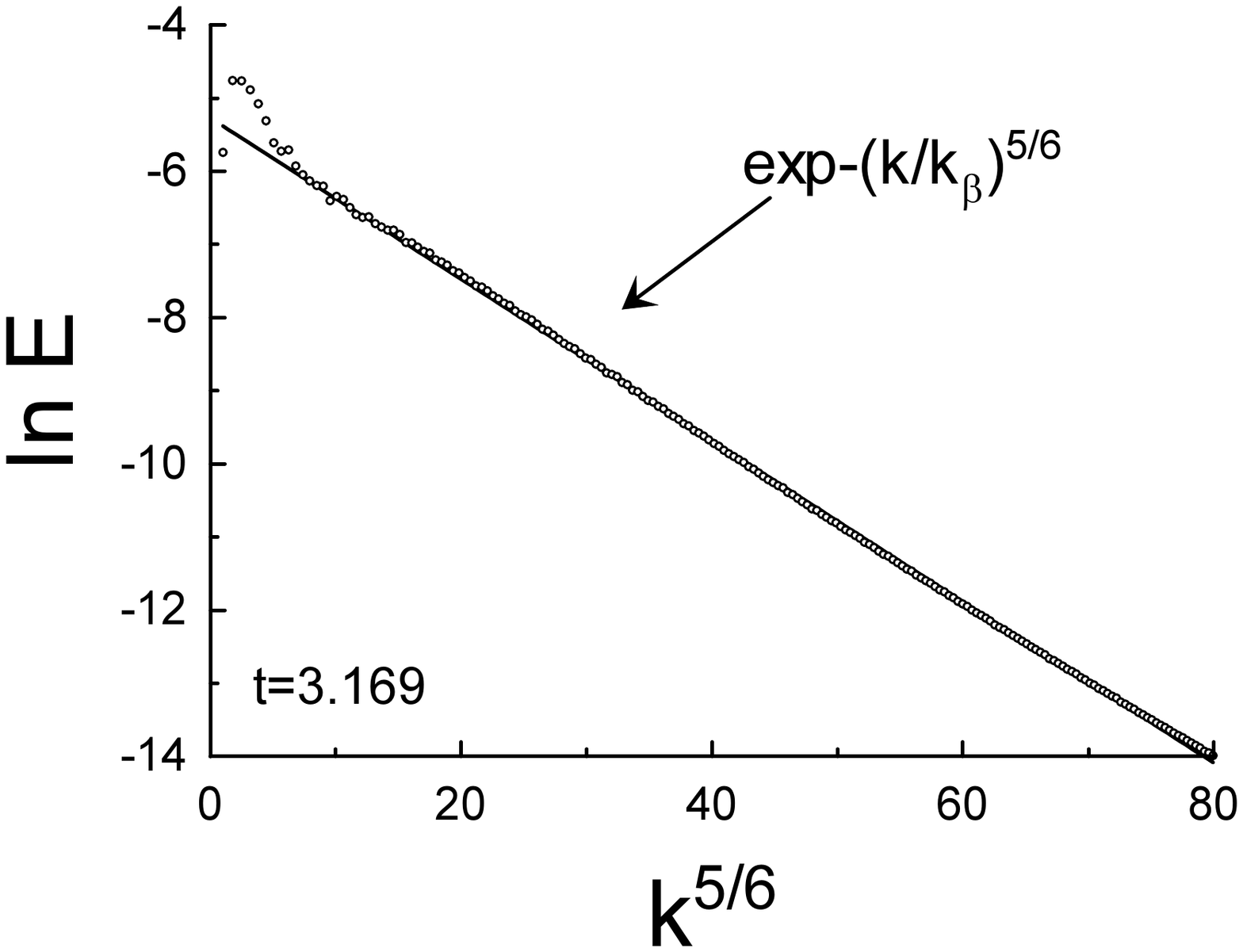}\vspace{-4cm}
\caption{\label{fig3} The same as in Fig. 2 but for $t=3.169$.}
\end{center}
\end{figure}

\begin{figure}
\begin{center}
\includegraphics[width=8cm \vspace{-0.1cm}]{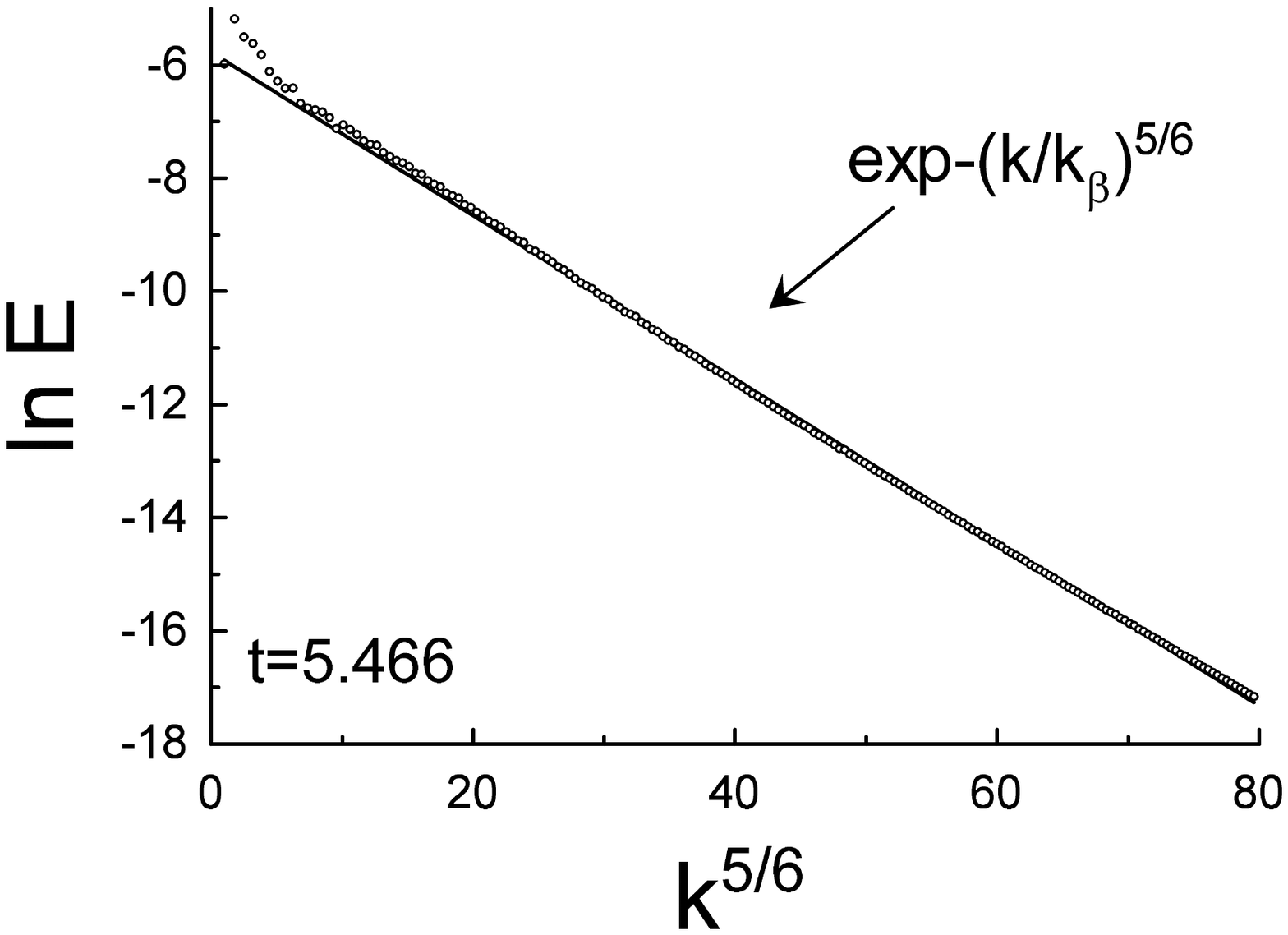}\vspace{-4cm}
\caption{\label{fig4} The same as in Fig. 2 but for $t=5.466$.}
\end{center}
\end{figure}
\begin{figure}
\begin{center}
\includegraphics[width=8cm \vspace{-1cm}]{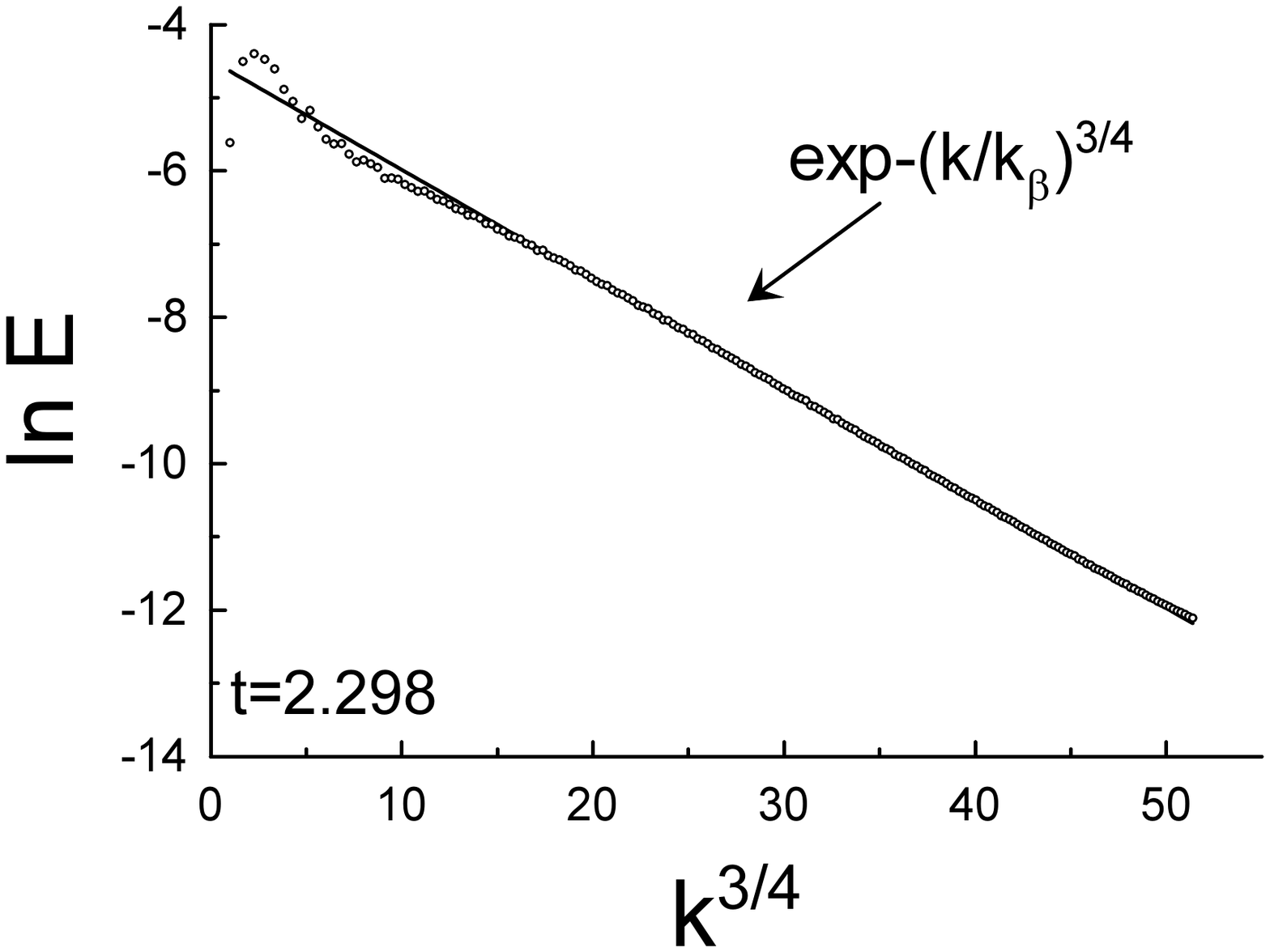}\vspace{-3.7cm}
\caption{\label{fig5} The same as in Fig. 2 but against  $k^{3/4}$.}
\end{center}
\end{figure}
The velocity power spectra for the distributed chaos 
have a stretched exponential form \cite{b1}
$$
E(k )   \sim \exp-(k/k_{\beta})^{\beta}  \eqno{(3)}
$$
In the asymptotic theory suggested in the Ref. \cite{b1} scale invariance of the group velocity of the waves driving the chaos
$$
\upsilon (\kappa ) \sim I_n^{1/2} \kappa^{\alpha_n}     \eqno{(4)}
$$
at $\kappa \rightarrow \infty $, provides $\alpha_n = (n+1)/2$ by the dimensional considerations,
and then
$$
\beta_n =\frac{2\alpha_n}{1+2\alpha_n}=\frac{n+1}{n+2}   \eqno{(5)}.
$$

\begin{figure}
\begin{center}
\includegraphics[width=8cm \vspace{-1cm}]{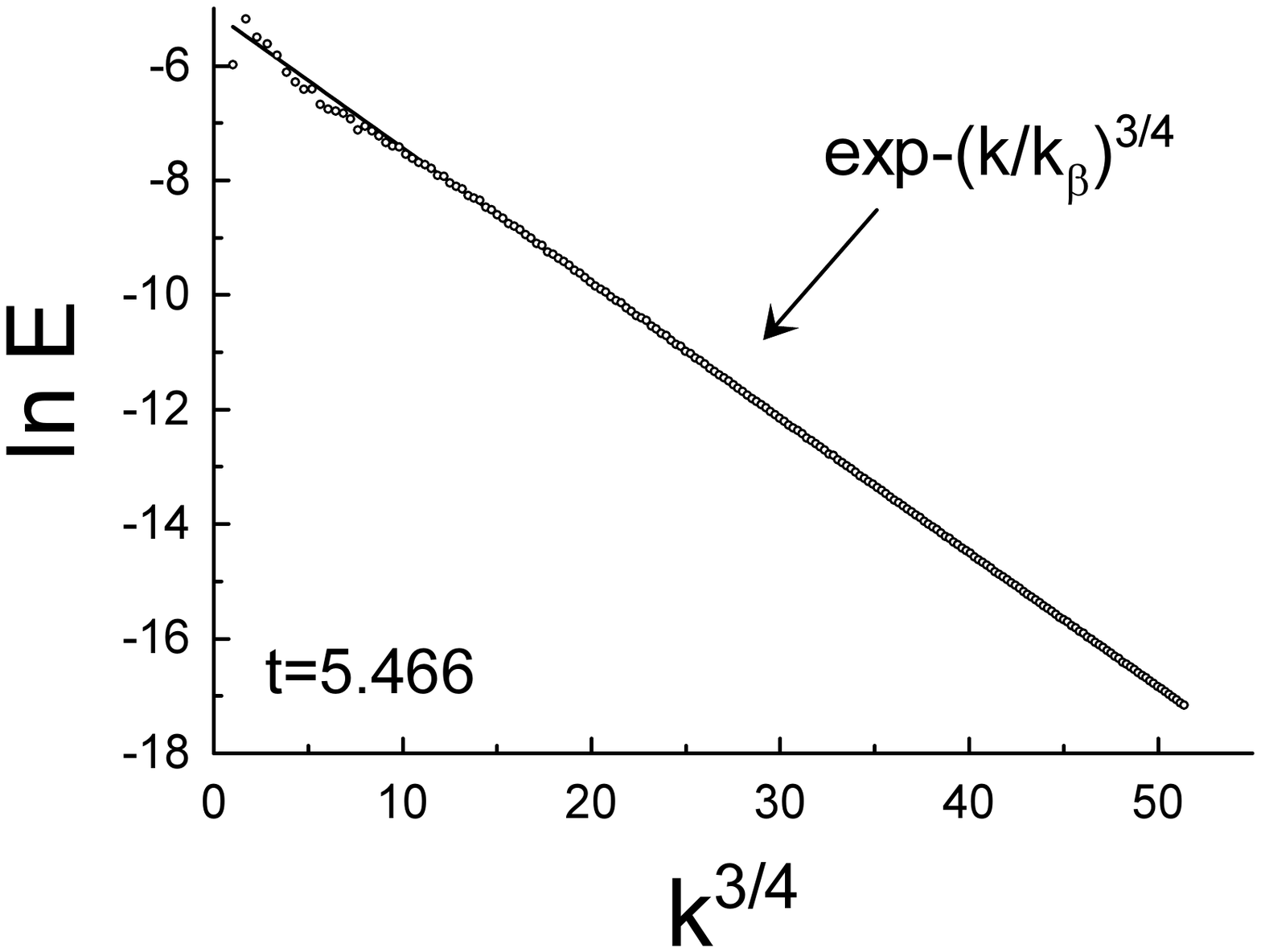}\vspace{-3.7cm}
\caption{\label{fig6} The same as in Fig. 4 but against  $k^{3/4}$.}
\end{center}
\end{figure}
For the h-attractor (Birkhoff-Saffman invariant) $\beta_2=3/4$ whereas for the i-attractor (Loitsyanskii invariant) $\beta_4=5/6$. Figures 2-4 show data of the same DNS \cite{torr},\cite{wr} as in Fig. 1 but for 3D energy spectra $E(k,t)$ of the free decaying isotropic turbulence at $t=2.298,~3.169,~5.466$. In the scales of the axes chosen for Figs. 2-4 the stretched exponential spectral law  Eq. (3) with $\beta=5/6$ corresponds to a straight line. One can see that $\beta_4=5/6$ provides good correspondence to the data for Figs. 2 and 3. 

  Figures 5 and 6 show the same data as in Figs. 2 and 4 in the scales suitable for the 
h-attractor value of $\beta_2=3/4$ (i.e. $\ln E$ against $k^{3/4}$), for comparison. One can see that competitiveness of the h-attractor against i-attractor increases with time, as expected for the developing attractors. Actually it seems from comparison of the Figs. 4 and 6 that at $t=5.466$ we are close to transition from the domination of i-attractor to domination of h-attractor, as in statistically stationary isotropic homogeneous turbulence \cite{b1} (cf. Fig. 1: the skewness in Fig. 1 saturates on the stationary turbulence value - the dashed straight line, and also Ref. \cite{p}).

\section{Acknowledgement}

I thank  A.A. Wray for sharing his data and J. Jimenez for organizing the AGARD site Ref. \cite{torr}.

\end{document}